\documentclass[aps,prl,reprint,superscriptaddress]{revtex4-1}

\usepackage{graphicx}% Include figure files

\begin{document}

\title{Ballistic transport in graphene antidot lattices}

\author{Andreas Sandner}
\author{Tobias Preis}
\author{Christian Schell}
\author{Paula Giudici }
\altaffiliation[present address: ]{Consejo Nacional de Investigaciones Cient\'\i ficas y T\'ecnicas (CONICET), Buenos Aires, Argentina}
\affiliation{Institute of Experimental and Applied Physics, University of Regensburg, D-93040 Regensburg, Germany}
\author{Kenji Watanabe}
\author{Takashi Taniguchi}
\affiliation{National Institute for Materials Science, 1-1 Namiki, Tsukuba 305-0044, Japan}
\author{Dieter Weiss}
\author{Jonathan Eroms}
\email[]{jonathan.eroms@ur.de}
\affiliation{Institute of Experimental and Applied Physics, University of Regensburg, D-93040 Regensburg, Germany}

\date{\today}

\begin{abstract}\bf

Graphene samples can have a very high carrier mobility if influences from the substrate and the environment are minimized. Embedding a graphene sheet into a heterostructure with hexagonal boron nitride (hBN) on both sides was shown to be a particularly efficient way of achieving a high bulk mobility \cite{Wang1DContacts}. Nanopatterning graphene can add extra damage and drastically reduce sample mobility by edge disorder \cite{PhysRevLett.102.056403,PhysRevB.82.041413,PhysRevB.85.195432}. Preparing etched graphene nanostructures on top of an hBN substrate instead of SiO$_2$ is no remedy, as transport characteristics are still dominated by edge roughness %due to the ion damage to the sample edges
\cite{BischoffBNRibbons}.
Here we show that etching fully encapsulated graphene on the nanoscale is more gentle and the high mobility can be preserved. To this end, we prepared graphene antidot lattices \cite{PhysRevLett.66.2790} where we observe magnetotransport features stemming from ballistic transport. Due to the short lattice period in our samples we can also explore the boundary between the classical and the quantum transport regime.

\end{abstract}

\pacs{}
%\keywords{}

\maketitle

In single layer graphene the charge carriers are completely exposed to the environment, which limits their mobility. Placing graphene on hexagonal boron nitride (hBN) was shown to improve the carrier mobility \cite{Dean2010}, allowing the observation of ballistic transport or the fractional quantum Hall effect in bulk graphene \cite{Dean2011}. Recently, a dry stacking technique was introduced, which allows complete encapsulation of graphene into layers of hBN and excludes any contamination from process chemicals such as electron beam resist \cite{Wang1DContacts}. To obtain graphene nanodevices, chemically prepared graphene nanostructures \cite{Jiao2009,Kosynkin2009,Cai2010} are a potential route for certain applications, however, the high flexibility of a top down patterning approach is extremely desirable.
Graphene antidot lattices can help circumventing the problem of the missing band gap in transistor applications \cite{Bai2010}, and were even predicted to serve as the technological basis for spin qubits \cite{PhysRevLett.100.136804}. Clearly, for advanced graphene nanodevices, not only the bulk mobility has to be improved, but the nanopatterning has to be optimised. 

Here we present experiments on graphene antidot lattices \cite{PhysRevLett.66.2790,Eroms2009,Shen2008} etched into hBN/graphene/hBN heterostructures with lattice periods going down to $a=50$ nm. Magnetotransport on those samples shows commensurability features stemming from ballistic orbits around one or several antidots. This allows us to prove that the high carrier mobility is preserved in the nanopatterning step even though the zero field resistance is dominated by scattering on the artificial nanopattern, giving an apparent reduction of the mobility. 
The small feature size of our samples also allows us to approach the region where the classical picture of cyclotron orbits no longer applies. This classical to quantum  crossover is governed by the ratio between the Fermi wavelength $\lambda_F$ of the carriers and the dimensions of the nanopattern.

To obtain embedded graphene samples, hBN/graphene/hBN stacks were prepared using the dry stacking technique, patterned into Hall bar shape, and contacted using Cr/Au \cite{Wang1DContacts}. In hBN/graphene/hBN samples prepared by this method, we routinely obtained carrier mobilities in excess of $\mu = 100\,000$ cm$^2$/Vs, showing all integer quantum Hall states starting from a few Tesla. In one sample without antidots and a mobility of $\mu = 300\,000$ cm$^2$/Vs we also observed the fractional quantum Hall effect at $T=1.4$ K. This shows that our fabrication procedure is mature and consistently yields high sample qualities. The samples presented in this study did not show any signs of a moir\'e superlattice \cite{Dean2013,GeimMoire}. Afterwards,  an antidot lattice was patterned. (For fabrication details, see Methods section). Fig. 1 shows an optical micrograph of a finished sample and a scanning electron micrograph of a sample after measuring. A sketch of the antidot lattice, etched into the stack is also shown.
The antidot lattice period $a$ was varied between 50 nm and 250 nm. The antidot diameter $d$ was lithographically defined to be about 40 nm, but due to the conical etching profile, the actual diameter in the graphene plane is smaller. Using SEM inspection, we estimate it to be about $25\dots 30$ nm.

In Fig. 2, we show data for a sample with a lattice period of $a=200$ nm. From the gate response of the conductivity at a magnetic field $B=0$, shown in Fig. 2a, we calculate an apparent field effect mobility of $\mu = 35\,000$ cm$^2$/Vs. At a carrier density $n_S = 2.3\times 10^{12}$ cm$^{-2}$ this corresponds to an apparent mean free path of about $l_{mfp}=\frac{\hbar}{e}\sqrt{\pi n_S}\mu=620$ nm. We estimate the intrinsic mean free path to be about 1400 nm  \cite{Ishizaka1999}(see Supplementary Information). 
Magnetotransport traces of this device (see Fig. 2b) show pronounced peaks at field values where the cyclotron diameter $2R_C = \frac{\hbar}{eB}\sqrt{\pi n_S}$ is commensurate to the square antidot lattice. The peak belonging to $2R_C = a$, the fundamental antidot peak, is most pronounced. Additional peaks appearing at lower fields correspond to orbits encircling 2, and 4 antidots \cite{PhysRevLett.66.2790} (see Fig. 1d), confirming a mean free path which spans several lattice periods. While in a simple picture only the unperturbed orbits encircling the antidots are responsible for the magnetotransport features, a more detailed analysis based on the Kubo formula shows that velocity correlations in the chaotic trajectories, which occupy the largest part of the phase space, result in the magnetoresistance peaks \cite{PhysRevLett.68.1367,PhysRevB.55.16331}. Most of the orbits therefore hit the antidot edges several times within a mean free path. Hence, the visibility of the antidot peaks not only proves a high bulk mobility, but also shows that scattering at the edges does not cut off the trajectories and we can conclude that the high carrier mobility also survives after nanopatterning.

At higher fields, the cyclotron diameter $2R_C$ is reduced below the neck width $a-d$ in between the antidots. We can observe Shubnikov-de Haas oscillations, eventually resulting in a well-defined quantum Hall effect. At $B=14$ T we clearly observe the $\nu=1$ plateau, which again shows the high sample quality (Fig. 2c).
We evaluated the carrier density dependence of the magnetoresistance peaks corresponding to orbits around 1, 2 and 4 (Fig. 2d) and found that the peaks were always well described by a square root dependence of the cyclotron diameter on the carrier density down to $n_S = 3.2\times 10^{11}$ cm$^{-2}$. Quantitatively, we confirmed the formula for the cyclotron diameter for graphene given above, which contains spin and valley degeneracy.

Fig. 3a shows the magnetoresistance of a sample with $a=100$ nm at $n_S = 2.8\times 10^{12}$ cm$^{-2}$. The apparent Hall mobility
at this density is about  $\mu = 8\,000$ cm$^2$/Vs. Again, scattering at the antidot potential limits the apparent mobility \cite{Ishizaka1999}, but the intrinsic mobility is higher as we clearly observe magnetoresistance peaks for $n=1, 2, 4$  antidots, and a fourth peak at lower fields is weakly visible. Ishizaka and Ando studied how the visibility of the higher order antidot peaks depends on the mobility \cite{PhysRevB.55.16331}. From their data, we estimate that the intrinsic mean free path must be at least 400 nm, well in excess of the apparent mean free path of 160 nm (see Supplementary Information).

The good visibility of the $n=2$ peak confirms the small aspect ratio $d/a \le 0.3$  \cite{PhysRevB.55.16331}, in agreement with our SEM analysis and also with the onset of the  well-defined Shubnikov-de Haas oscillations in our magnetotransport data. All these approaches give an antidot diameter of $d=25\dots 30$ nm. 
 
In experiments in GaAs based antidot lattices it was found that due to depletion at the antidot boundaries, the potential can be very soft and small lattice periods are hard to realize. In our case the data compares well to hard-wall potential lattices in GaAs, which could be realized in GaAs only at much larger lattice periods \cite{PhysRevLett.66.2790}. We also compared data for similar carrier densities in the electron and hole regime in Fig. 3b and found the graphs to be virtually identical. This proves that there is no edge doping at the antidot boundaries, which would have led to different potential shapes in the electron and hole regime due to Fermi level pinning at the edges.

Now let us discuss the transition between the quantum and the classical transport regime. 
In GaAs-based heterostructures, the smallest lattice period realised so far was $a=80$ nm, and required critical tuning of the etch depth \cite{Antidots80nm}. In contrast, due to the lack of a depletion region in graphene the fabrication of samples with a very small lattice period is less critical, and the carrier density is widely tunable. Also, due to valley degeneracy, the Fermi wavelength in graphene, $\lambda_F = 2\sqrt{\frac{\pi}{ n_S}}$ is a factor of $\sqrt{2}$ larger than in GaAs based 2DEGs at the same carrier density. Thus we can explore the transition from the semi-classical to the quantum regime \cite{Brack1997}, where a description in terms of classical orbits is no longer justified. In the samples with $a\le 100$ nm we are able to study this transition. Fig. 4a shows the disappearance of the main antidot peak in a sample with $a=75$ nm as the carrier density is lowered, making $\lambda_F$ longer. We find that this peak is only visible at densities above $n_S = 4.3\times 10^{11}$ cm$^{-2}$, corresponding to $\lambda_F= 54$ nm. Also, in two samples with $a=100$ nm, we observe that the main antidot peak becomes visible for densities larger than $n_S = 2.2\times 10^{11}$ cm$^{-2}$, which corresponds to  $\lambda_F = 75$ nm. In a sample with $a = 50$ nm, we observed a weak antidot peak only at $n_S = 2.5\times 10^{12}$ cm$^{-2}$($\lambda_F= 22$ nm).
To be in the classical limit of a quantum system, the Fermi wavelength must satisfy a condition $\frac{\lambda_F}{2\pi}
 \ll l$ \cite{SakuraiQM}, where $l$ is a typical dimension of the system. In our case, the neck width $a-d$ of the constriction between the antidots is the shortest length scale in the problem, and we find that when $\lambda_F \approx a-d$ the classical regime sets in and the antidot peak becomes visible. 

The fact that the antidot peaks disappear at low densities can be either due to a limited mean free path or the breakdown of the classical picture.  In Fig. 2d (lattice period $a=200$ nm) all the antidot peaks disappear at roughly the same magnetic field, $B\approx 0.5$ T (where $\mu B$ exceeds some constant), but different carrier density. This behaviour is clearly governed by a limited mean free path.
In contrast, in the sample of Fig. 4a ($a=100$ nm), we find that the classical features at both $B\approx 1$ T and $B\approx 2.5$ T disappear at the same carrier densities, making a $\lambda_F$-driven scenario more realistic.

Finally, at low densities, we can observe a weak localization (WL) feature at low temperatures: a peak in the magnetoresistance at $B=0$ (see Fig. 4b). Using a standard analysis for WL in graphene \cite{PhysRevLett.97.146805} that we employed in earlier work on graphene antidot lattices on SiO$_2$ \cite{Eroms2009}, we extracted the phase coherence length $L_\phi$. For the sample with $a=100$ nm (same as in Fig. 3a) we found it to be between 120 nm and 300 nm (see Fig. 4c). It clearly exceeds the lattice period, unlike in graphene antidot samples on SiO$_2$ where $L_\phi$ was significantly below $a$ \cite{Eroms2009}. We therefore again conclude that nanopatterning of embedded graphene leads to greatly reduced scattering at the sample edges.

In summary, we prepared antidot lattices in stacks of hBN/graphene/hBN and observed well-developed commensurability features in samples with lattice periods from $a=50$ nm to $a=250$ nm. This shows that the etching procedure preserves the high sample quality. In the short-period graphene samples, we could observe the disappearance of classical features when the Fermi wavelength $\lambda_F$ exceeds $a-d$, marking a classical to quantum transition. Our experiments therefore pave the way for well-controlled graphene based nanodevices.

\section{Methods}

Single crystalline hexagonal boron nitride (hBN)\cite{Kubota2007,Taniguchi2007} was exfoliated onto a stack of PMGI and PMMA polymers spin coated on an oxidised Si-Wafer \cite{Dean2010}. Suitable hBN flakes, serving later as the top hBN layer, were located in an optical microscope by using different bandpass filters. The PMGI sacrificial layer was dissolved in photoresist developer and DI water, leaving the PMMA with the hBN floating on DI water. Then the PMMA film was transferred to a microscope slide into which a hole had been cut \cite{Dean2010}. Single layer graphene was exfoliated from HOPG (Momentive Performance, ZYA grade) onto oxidised Si, and picked up by van-der-Waals interaction using the first hBN flake \cite{Wang1DContacts}. Using a home-made setup in an optical microscope, this stack was transferred to a second hBN flake, residing on an oxidised Si substrate with prepatterned markers. 
Subsequently, the finished stacks were annealed in forming gas flow at 320$^\circ\text{C}$ for several hours.
The hBN/graphene/hBN heterostructure was then patterned into Hall bar shape using electron beam lithography (EBL) and CHF$_3$/O$_2$ (40 sccm/6 sccm, 60 Watt power) based reactive ion etching (RIE) \cite{Wang1DContacts}. Cr/Au side contacts were defined with EBL and deposited by thermal evaporation and lift-off after brief oxygen plasma cleaning of the contact areas \cite{Wang1DContacts}. Finally, the antidot lattice was defined in a separate EBL and RIE step.
Samples were glued into chip carriers with silver filled epoxy to contact the back gate, wire-bonded and measured in a helium cryostat with variable temperature insert, using low frequency lock in techniques with a bias current of 10 nA.

\section{Acknowledgments}

The authors thank A. Geim and R. Jalil for sharing details of the graphene transfer procedure, R. Fleischmann, T. Geisel and K. Richter for helpful discussions, and the Deutsche Forschungsgemeinschaft (DFG) for funding through projects GRK 1570 and GI 539/4-1.

%\section{Author Contributions}

%A.S. performed the sample fabrication and measurements, A.S., T.P., C. S. and P.G. developed the graphene transfer procedure, K.W. and T.T. grew the hBN crystals, J.E. designed the experiment, J.E., A.S. and D.W. analysed the data and co-wrote the manuscript, all authors edited the manuscript.
%\section{References}
%\bibliographystyle{apsrev4-1}
%\bibliographystyle{naturemag_noURL}
%\bibliography{Antidots_Graphene}

\cleardoublepage
\section{Figures}

\begin{figure}%
\includegraphics[width=8.3cm]{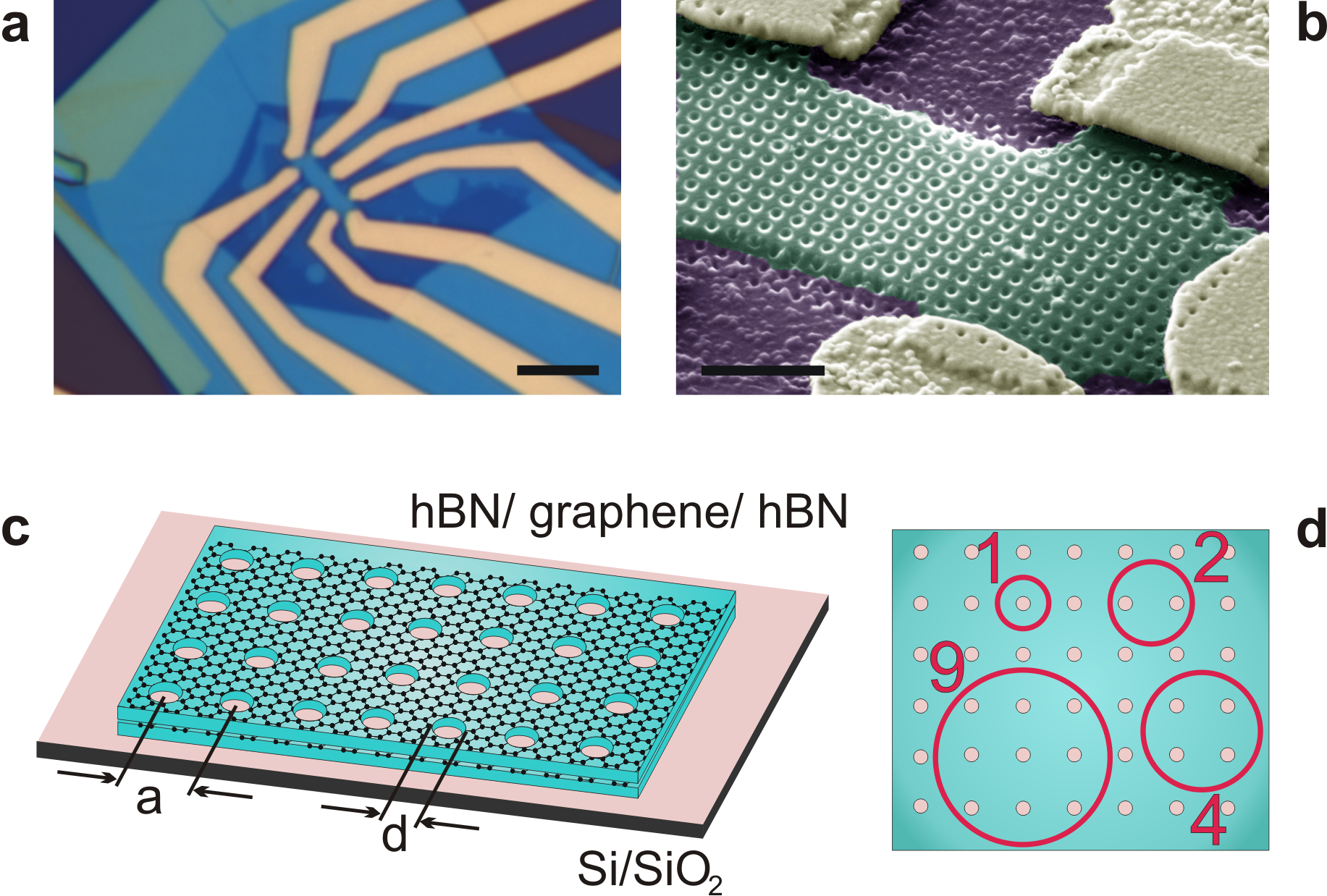}%
\caption{\textbf{a}, Optical micrograph of a finished graphene Hall bar etched out of an hBN/graphene/hBN heterostructure, and contacted with Cr/Au leads. Scale bar length 5 $\mu$m. \textbf{b}, False-color scanning electron micrograph of a sample with lattice period $a=100$ nm. Here, the heterostructure is shaded in green, the Cr/Au contacts yellow, and the Si/SiO$_2$ substrate violet. Scale bar length 500 nm. \textbf{c}, Sketch of the antidot lattice in an hBN/graphene/hBN heterostructure. The antidot lattice period $a$ ranges from 50 to 250 nm, and the antidot diameter $d$ is about $25\dots 30$ nm. \textbf{d}, The most prominent cyclotron orbits fitting into the lattice, giving rise to magnetoresistance peaks.}%
\label{Fig:SEM}%
\end{figure}

\begin{figure}%
\includegraphics[width=1\textwidth]{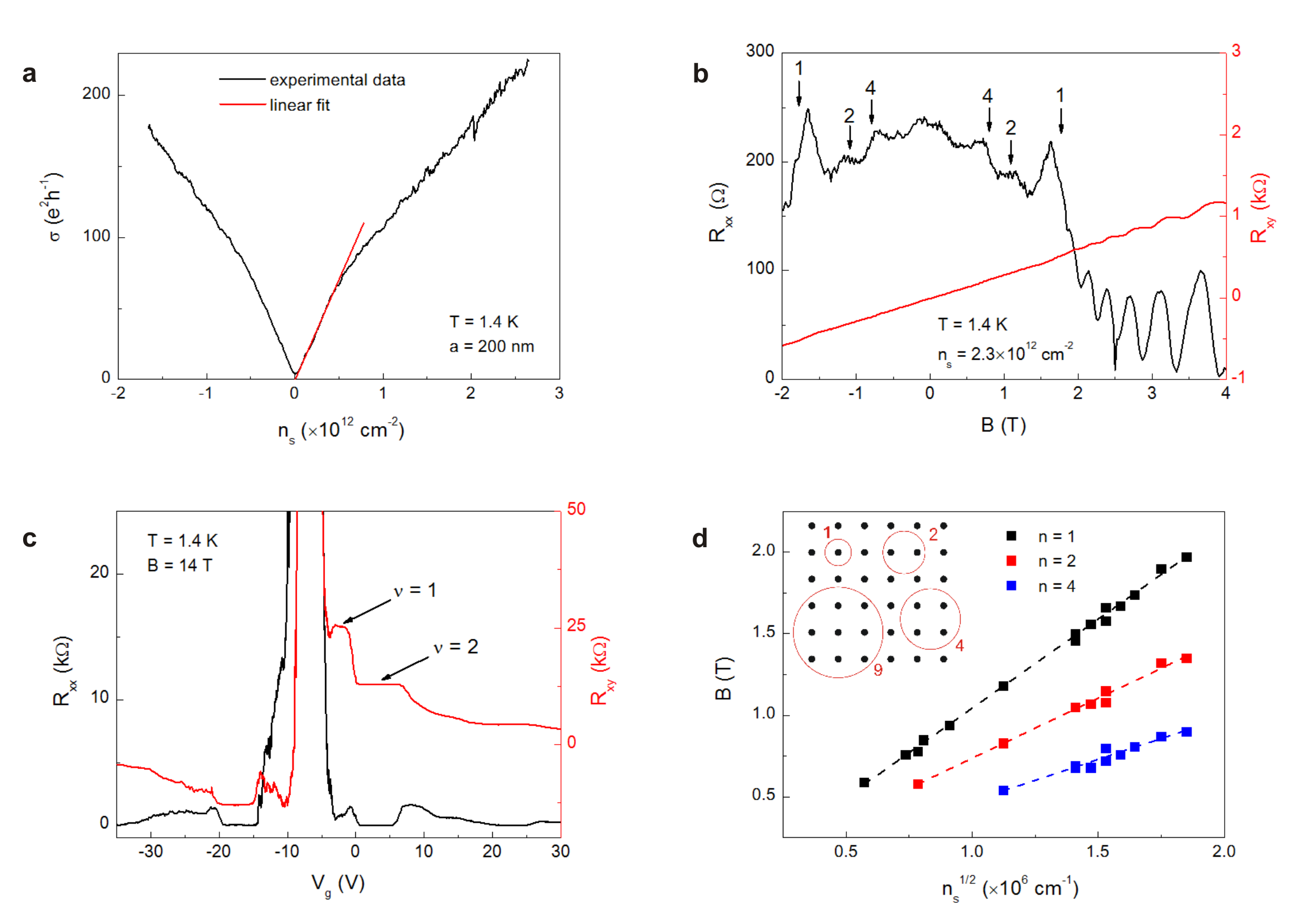}%
\caption{\textbf{a}, Gate dependence of the sheet conductivity of a sample with $a=200$ nm. The linear fit (red line) gives an apparent mobility of $\mu = 35\,000$ cm$^2$/Vs. \textbf{b}, Magnetoresistance (black) and Hall resistance (red). The arrows correspond to the expected magnetic field positions of the orbits sketched in Fig 1d. The fine structure in $R_{xx}$ is not noise, but phase-coherent oscillations \cite{Nihey1993,PhysRevLett.70.4118,PhysRevB.49.8510} that disappear at higher temperatures. \textbf{c}, Gate dependence of $R_{xx}$ and $R_{xy}$ at $B=14$ T, showing a clear $\nu=1$ quantum Hall plateau. This feature is only observed in high-mobility graphene devices. \textbf{d}, The magnetic field positions of the three antidot peaks scale with the square root of the carrier density, confirming the classical nature of those peaks.}%
\label{Fig:200nm}%
\end{figure}

\begin{figure}%
\includegraphics[width=1\textwidth]{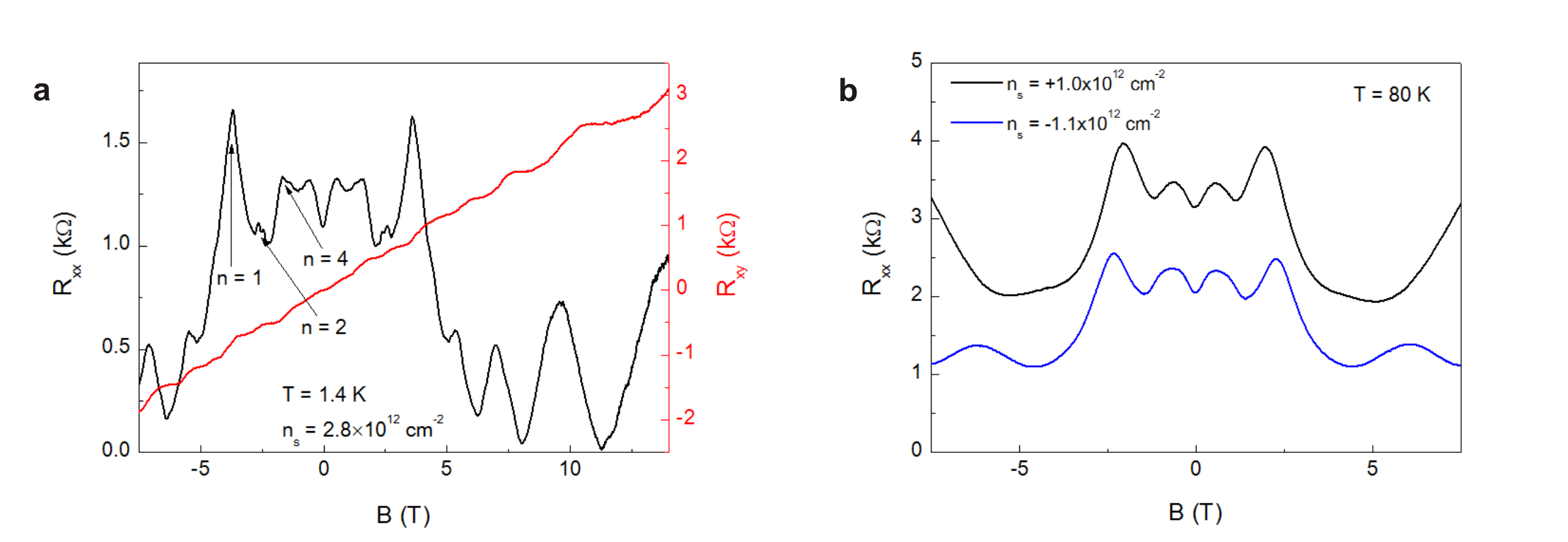}%
\caption{\textbf{a}, Magnetoresistance and Hall resistance data taken on a sample with $a=100$ nm. The three well-defined antidot peaks correspond to orbits around 1, 2 and 4 antidots. \textbf{b}, $R_{xx}$ taken at similar electron and hole density. Here, $T=80$ K to show the classical features more clearly. There is virtually no difference between those graphs, proving that the potential profile is the same for electrons and holes.}%
\label{Fig:100nm}%
\end{figure}

\begin{figure}%
\includegraphics[width=1\textwidth]{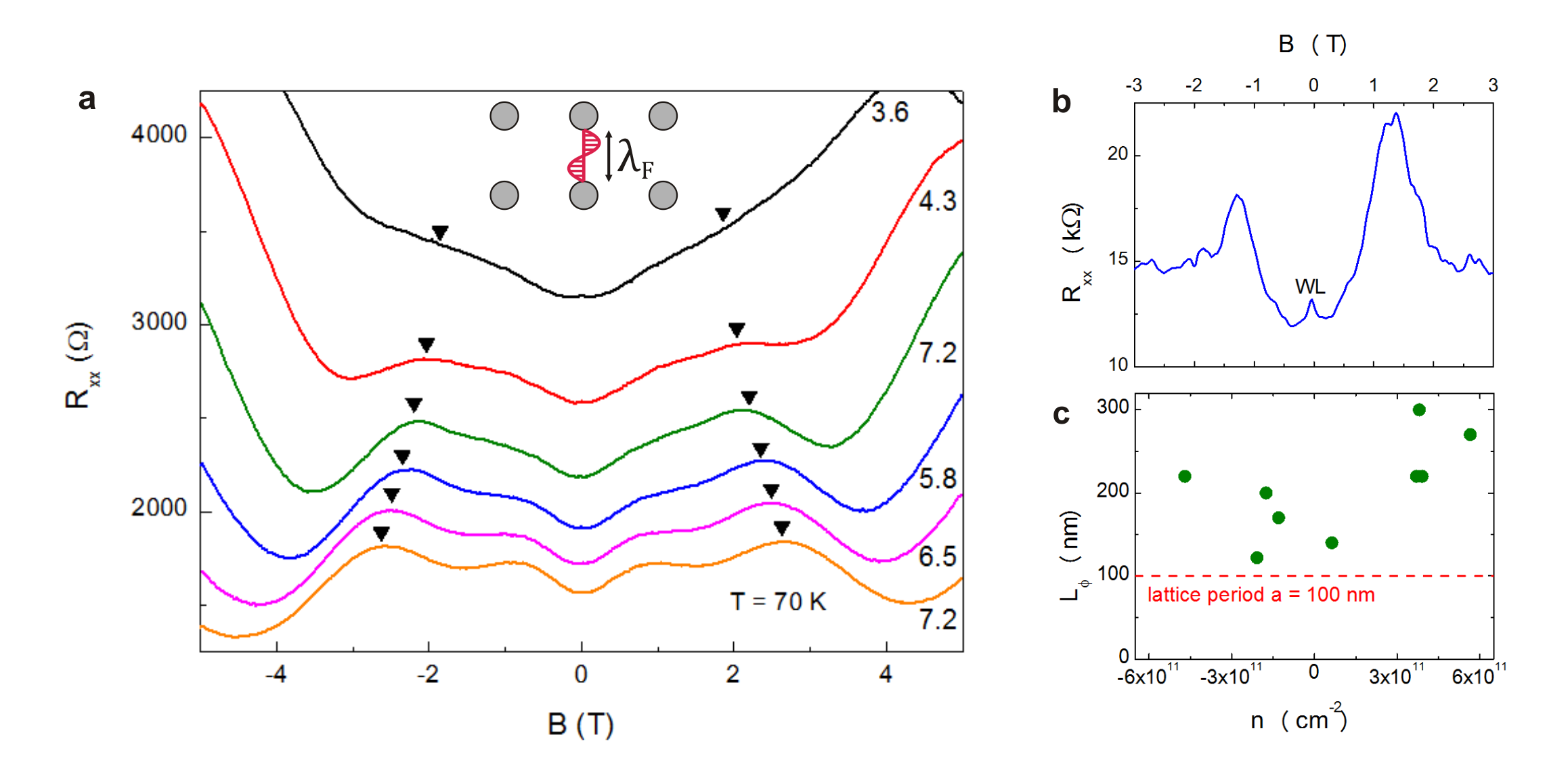}%
\caption{\textbf{a}, $R_{xx}$ data of a sample with $a=75$ nm taken at very low densities, at the transition into the regime of classical transport. %Charge neutrality point was at $V_{g}=-1$ V. Triangles show the expected position of the main antidot peak.
The densities $n_S$ are given in units of $10^{11}$ cm$^{-2}$, shown next to the corresponding graphs. The expected position of the main antidot peak is marked with a triangle for each density. A higher order antidot peak is also visible at lower magnetic fields. As the carrier density is lowered, the antidot peaks disappear. Inset: Sketch of the Fermi wavelength corresponding to the red graph. \textbf{b}, Weak localization (WL) peak in the sample with $a=100$ nm, taken at $n_S=1.3\times 10^{11}$ cm$^{-2}$. The antidot peak is not visible at this low density, the big peaks at $B=\pm 1.4$ T are a Shubnikov-de Haas oscillation. \textbf{c}, Phase coherence length taken from the weak localization fits of a sample with $a=100$ nm at various low densities and $T=1.4$ K. The phase coherence length exceeds the lattice period, showing that the etched boundaries do not lead to severe phase-breaking.}%
\label{Fig:lowdens}%
\end{figure}

\end{document}